# Analyzing the Kasiski Method Against Vigenere Cipher

April Lia Hananto [1], Arip Solehudin [2], Agung Susilo Yuda Irawan[3], Bayu Priyatna [4]

[1,4]Faculty of Technology and Computer Science, Universitas Buana Perjuangan Karawang, Indonesia
aprilia@ubpkarawang.ac.id, bayu.priyatna@ubpkarawang.ac.id

[1]* Faculty of Computing, Universti Teknologi Malaysia Skudai Johor, Malaysia
hananto1983@graduate.utm.my

[2, 3]Faculty of Computer Science, Universitas Singaperbangsa Karawang, Indonesia
arip.solehudin@staff.unsika.ac.id , agung@unsika.ac.id

------------------------------------************************---------------------------------

## Abstract:

The weakness of the vigenere cipher lies in its short key and is repeated, so there is a key loop in encrypting messages, this is used by cryptanalysts using the Kasiski method to know the key length so it can solve this algorithm. The Kasiski method uses repetitive cryptograms found in the ciphertext to determine the key length. Modification of the vigenere cipher solves strengthen the cipher by using arranged keys to make it difficult to crack the keys against the Kasiski method attacks. This study analyzes the strength of the vigenere cipher ciphertext and the modification of the vigenere cipher from the encryption results composed of plaintexts that use different keys. Against the Kasiski method attack. The method used in this study is a quantitative method with a descriptive approach through nonparametric statistical tests on the sign test. The results showed a difference in the ciphertext's strength from the encryption process even though using the same key. The kaiseki method if can know not all key lengths the ciphertext in the compilation does not occur repetitive cryptograms. Modification of the vigenere cipher has a positive effect on the strength of the cipher and can influence the strength of the ciphertext that is built against the Kasiski method attacks.

*Keywords* — **Vigenere Cipher, Kasiski Method, Quantitative Method.**

------------------------------------************************---------------------------------

## I. INTRODUCTION

Vigenere cipher is very well known because it is easy to understand and implement, but the weakness of the vigenere cipher lies in its short key and is repeated, so there is a key loop used to encrypt the plaintext. This is used by cryptanalysts using the Kasiski method, to know the length of the key used so it can solve this algorithm.

Modification of the vigenere cipher solves strengthen the cipher by using a arranged key with a key length equal to the length of the plaintext. I expect the purpose of this modification to strengthen the cipher and to complicate the key breakdown of the Kasiski method attack.

the amount of work can measure the level and strength of a cipher needed to break the ciphertext data into plaintext without knowing the key used in the encryption process. The more work needed, the longer it takes to break the ciphertext into plaintext without knowing the key to a cipher. This means that the stronger the cipher is, the more secure the cipher is used to protect the message to be kept confidential. This study describes how the strength of the cipher in the vigenere cipher algorithm and the modification of the vigenere cipher to the Kasiski method attack by knowing the key length used in the ciphertext (cipher attack).

## II. THEORETICAL BASIS

### A. *Cryptography*

Cryptography (cryptography) comes from Greek, crypto and graphic. Crypto means secret, while graphic means writing [4]. Cryptography is the study of mathematical techniques related to aspects of information security aimed at maintaining message confidentiality, data integrity, and authentication.

Cryptography techniques have long been believed to handle message security issues, because, besides using computer programming languages, cryptography uses mathematical formulas and notations in mathematics, ranging from simple formulas to complex formulas which include theories, theorems and definitions in mathematics [6]. Cryptography (cryptography) in Greek is divided into two terms namely "cryptós" which has a secret meaning, while "gráphein" means writing, from both terms combined to become "secret writing" [1].

Cryptography is the science and art of protecting messages when messages are sent from one place to another [3]. Cryptographic techniques in changing messages that can be read or original text (plaintext) into certain codes or ciphertext messages are called encryption [7].

The algorithm used in the encryption and decryption process is a cryptographic algorithm called a cipher. They often refer the term ciphertext to as a cryptogram. Encryption and decryption operations can be modeled:

EK (P) = C (Encryption Process)
DK (C) = P (Decryption Process)

At the time of the E encryption process, the message on the P plaintext by using a K key will produce a ciphertext message C. While the decryption process D, the Ciphertext C message deciphered by using the K key, will return the same P plaintext message as the previous message.

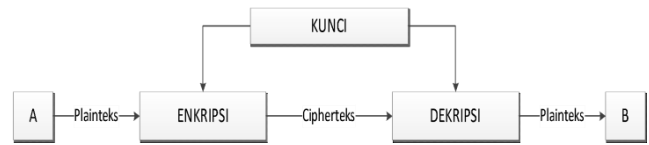

Fig 1. Encryption and Decryption Process

### B. *Vigenere Cipher*

Vigenere cipher is a substitution technique of polyalphabetic substitution. This cipher is easy to understand and implement, therefore the vigenere cipher is very well known. Although (with the help of computers), many computer security programs that use this cipher [6]. The encryption process is carried out by this cipher, by substituting plaintext letters for the characters (letters or numbers) of the key used and vice versa the decryption process substitutes the ciphertext letters with the same key (symmetric key). If the key length is shorter than the length of the plaintext, the key will be repeated, the mathematical model of the vigenere cipher encryption and decryption process is:

$Ci = (Pi + Kr) \mod 26$ (encryption process)
$Pi = (Ci - Kr) \mod 26$ (decryption process)

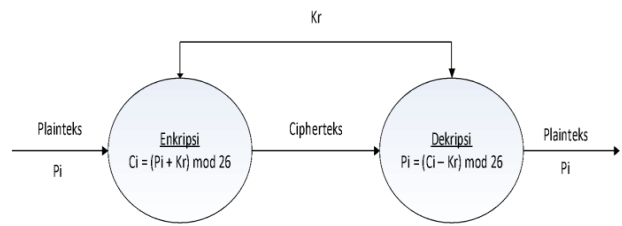

Fig 2. Encryption of the Vigenere Cipher Decryption

With Ci modeling the i-th letter in ciphertext, Pi modeling the i-th letter in plaintext, Kr models the r-the letter in the key. The above model is treated if A = 0, B = 1 to Z = 25 and mod 26 because the number of letters of the alphabet from A to Z is 26.

| 0 | 1 | 2 | 3 | 4 | 5 | 6 | 7 | 8 | 9 | 10 | 11 | 12 | 13 | 14 | 15 | 16 | 17 | 18 | 19 | 20 | 21 | 22 | 23 |
|---|---|---|---|---|---|---|---|---|---|----|----|----|----|----|----|----|----|----|----|----|----|----|----|
| A | B | C | D | E | F | G | H | I | J | K  | L  | M  | N  | O  | P  | Q  | R  | S  | T  | U  | V  | W  | X  |

Fig 3. Message Character Model

For example in applying vigenere encryption, if the plaintext is written "CYPTO IS SHORT FOR CRYPTOGRAPHY" and the key used is "ABCD" then the ciphertext formed is as follows:

Plaintext:
CRYPTOISSHORTFORCRYPTOGRAPHY
Key:
ABCDABCDABCDABCDABCDABCDABCD
Ciphertext:
CSASTPKVSIQUTGQUCSASTPIUAQJB

From the example above can be calculated through the mathematical model vigenere, for example in the letters Plaintext C and R with key letters A and B, namely:

*(C + A) mod 26 = (2 + 0) mod 26 = 2 = C (R + B) mod 26 = (17 + 1) mod 26 = 18 = S*

The decryption process in the above calculation on ciphertext C and S with keys A and B is as follows:

*(C - A) mod 26 = (2 - 0) mod 26 = 2 = C (SB) mod 26 = (18 - 1) mod 26 = 17 = R*

In the example above the key used is "ABCD", the character of the plaintext letters is longer than the key letters so that the key repeats periodically according to the length of the plaintext. This is the weakness of the vigenere cipher against the cases method when recurring cryptograms occur in the ciphertext.

### *C. Kasiski Method*

Method of solving the vigenere cipher, but the first person to publish the solution of the vigenere cipher was Friedrich Kasiski so this method was called the kaiseki method. The main disadvantages of the vigenere code are the short key and its repeated use. If the cryptanalyst can determine the length of the key to the repeating cryptogram, the ciphertext produces a key length possible to decrypt the ciphertext itself [6].

The Kasiski method helps find the length of the key by using advantages such as the English plaintext which contains not only letter repetition but also looping of pairs of letters or triple letters such as "TH", "THE", and so on. This letter repetition makes it possible to produce repetitive cryptograms [3].

The steps for the Kasiski method are:
a. Find all the repeating cryptograms inside the coded text (long messages contain repetitive cryptograms).
b. Calculate the distance between repeating cryptograms.
c. Calculate all the factors (divisors) of the distance (the divisor factor expresses the key length).
d. Determine the slices of the set of dividing factors. The value that appears in the slice represents the number that appears on all the dividing factors of the distances. This value may be the key length. This is because repeated strings can appear overlapping.

As an example:

**Plaintext:**
CRYPTOISSHORTFORCRYPTOGRAPHY

**Key:**
ABCDABCDABCDABCDABCDABCDABCD

**Ciphertext:**
CSASTPKVSIQUTGQUCSASTPIUAQJB

In this example, they encrypt the CRYPTO into the same cryptogram, which is CSASTP. This is because the distance between the two strings that repeat themselves in the plaintext is multiple of the key length so that the same string will appear to be the same cryptogram in the ciphertext. The distance between these two occurrences (calculated from the beginning of each string repeating) is 16 characters (not including spaces) and 16 is a multiple of 4. This is appropriate as shown in the example, where the key is arranged in length 4, "ABCD". The purpose of the Kasiski method is to find two or more repetitive cryptograms to determine the length of the key.

### *D. Test Sign*

The sign test is used to test the comparative hypothesis of two samples that are correlated if the data is in the form of ordinal. This technique is called the sign test because we express the data to be analyzedwe express the data to be analyzed in the form of signs, positive and negative signs. For example in an experiment, they do not state the results how much the change is but expressed in the form of positive and negative changes. For example, does the incentive given to employees have a positive effect on the effectiveness of a company? So, in this case, it is not stated how much influence, but only the statement has a positive or negative effect. The samples used in the sign test are samples are in pairs, for example, husband and wife, malefemale, public-private employees and others. I will know positive and negative signs based on differences in values between one another in the pair. As an illustration, for example: the null hypothesis (Ho) tested is: p (X> Y) = p (X <Y) = 0.5. The opportunity to change from X to Y is the same as the chance to change from Y to X = 0.5 or the chance to get a positive difference is the same as the chance to get a negative difference. So if the positive sign is far more than the negative sign, then Ho is rejected. With X = value after treatment (treatment) and Y = value before treatment. I can also know Ho based on the median of the group observed [11].

### *E. DescriptiveMethod*

Descriptive research is research aimed at describing or describing existing phenomena, both natural phenomena and human engineering. These phenomena can be in the form of forms, activities, characteristics, changes, relationships, similarities, and differences in other phenomena [13].

That descriptive research is a research method that seeks to describe and interpret objects according to what they are and researchers do not control and manipulate research variables. Therefore this research is called non-experimental research [12].

## III. RESEARCH METHOD

The method used in this study is a quantitative research method with a descriptive approach through nonparametric statistical tests on the sign test. This research will test the strength of the cipher and analyze the data on the encryption results that are composed of plaintexts that use different keys in the two algorithms (vigenere cipher and modification of the vigenere cipher) against the Kasiski method attack using the cipher attack technique. The object of this research is the strength of the cipher on the genre algorithm and the modification of vigenere to the Kasiski method attack. The parameter used is the ordinal value of the strength (strong or weak) of the vigenere cipher of each key used in the predefined plaintexts. Three types of key indicators are parameters, namely the short key, medium key, and long key. Each key that is built will be measured the strength and time needed to attack the Kasiski method using a standard algorithm (vigenere cipher) and modification algorithm (vigenere cipher).

The quantitative research process that can be developed in a flowchart is as follows:

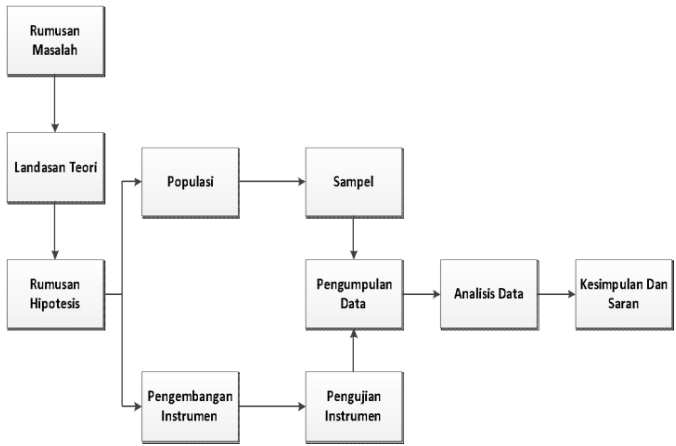

Fig 4. Quantitative Research Process [10]

### A. Population and Sample

Population (N) in this study is a collection of ciphertext built by the keys used in the encryption process in predefined plaintexts. While the sample (n) in this study is a ciphertext formed by keys composed of short, medium and long keys. There are six text sheets in the message with three messages written in English and three messages in Indonesian. Each plaintext will be formed by ten types of keys, including four short keys (4 to 6 letter characters), four intermediate keys (8 to 15 letter characters) and two long keys (11 to 25 letter characters) that are overall formed from five keys using Indonesian and five keys for using English. The results of the ciphertext are arranged in each algorithm (vigenere cipher and modification of vigenere cipher), with N = 60 and the sample was taken is the whole of the existing population. The number of samples is the total population (n = N).

### B. Data Collection Techniques

Data collection techniques used in this study are observation techniques (observation) on sample data taken through parameters (strong or weak ciphers) against the encryption arrangement of the ciphertexts formed through short, medium and long keys. Cipher algorithm is strong if there are no repeating cryptograms, whereas ciphers are weak when at least two or more cryptograms are repeated.

### C. Data Analysis Techniques

Data analysis techniques in this study used nonparametric statistical tests through the sign test in the SPSS application. Based on the name of the test the sign states the + sign and the sign - which is got from the difference in the observation's value to find out the effect of something. The sign test is used to compare the effect of the results of two treatments based on positive or negative signs of the difference between the observation pairs, treatment A (vigenere cipher) and treatment B (modification of vigenere cipher). The sign (+) shows there is no cryptogram for the two algorithms, while the sign (-) states there is a cryptogram for one algorithm.

### IV. RESULTS AND DISCUSSION

Algorithms and Vigenere Cipher Applications The application modification of the genre cipher is an application of the development of the previous vigenere cipher algorithm, while the difference in this modification lies in the repetition of the keys that were constructed as discussed in the previous chapter. We made application modification of the vigenere cipher to produce ciphertext from the encryption process and vice versa the decryption process to produce the original plaintext on the same key.

Encryption and decryption flowcharts in the changed Vigenere cipher application are:

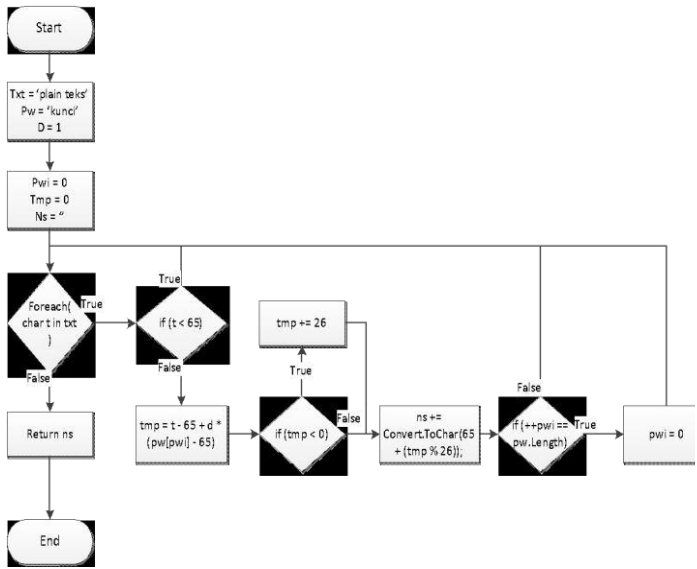

Fig 5. Changed Vigenere Cipher Flowchart

Image display application modification vigenere cipher is:

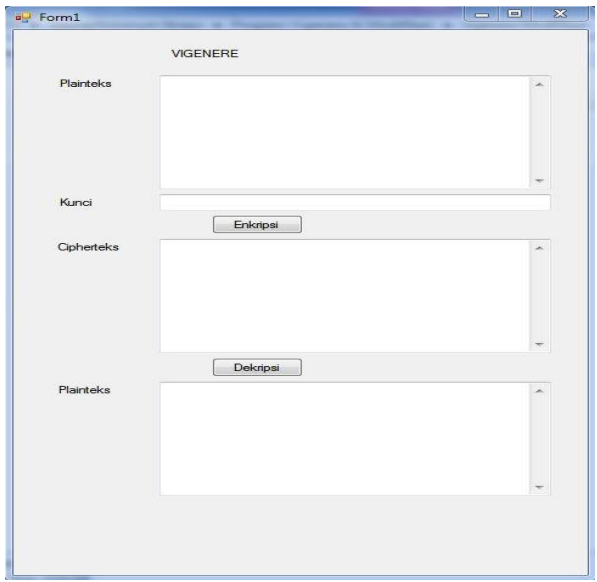

Fig 6. Vigenere Cipher Application

Like the previous vigenere cipher application, the vigenere cipher modification application is built together using a 26 character vigenere with a key limit that can be received a maximum of 256 characters. The main appearance in the application of the modification of the vigenere cipher in this study has four parts, namely plaintext, key, ciphertext and process (action) namely encryption and description.

### A. Kasiski Ciphertext Vigenere Cipher Method

In encrypting the Kasiski method of the vigenere cipher, it is by finding a cryptogram in a repetitive ciphertext due to periodic key repetition of the length of the plaintext which is arranged longer in encryption. In this study, the crypto or test tools through the Kasiski method to know the key length of the arranged ciphertext without having to know the key used.

**Plaintext:**
UNSIKA IS THE EXTENSION OF SINGAPER NATION KARAWANG UNIVERSITY

**Key:** ABCD (4 characters)

**Ciphertext:**
UOULKBCGAMCKKFRDNKCQGBPGASKXNJ XHRTKWATULNHCSESDDNHUDKBTDWBPJ

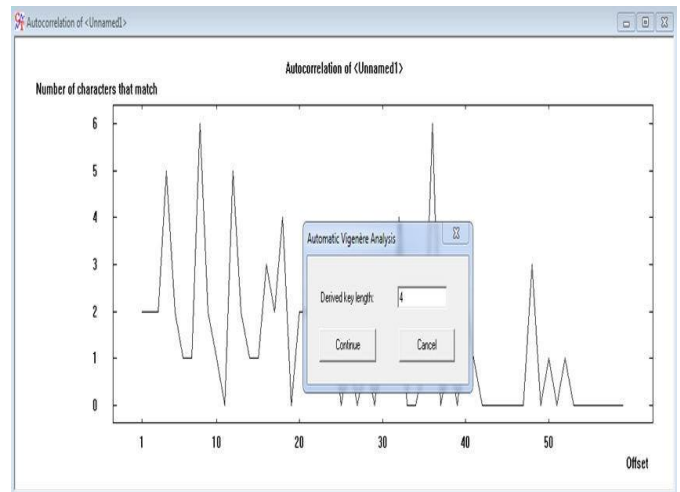

Fig 7. Kasiski on the Cryptool Application

### B. Determination of Plaintext

Several plaintexts that are only used in the * .txt file format will be presented, including three plaintexts in English and three plaintexts in Indonesian.

## C. Key Formation

The key formation in this study is based on the length of a key, it is said to be a long key if the key length is more than 128 bits, whereas the key is short if the key length is less than 56 bits and the key is the medium length between 56 bits and 128 bits.

## D. Test Results

Testing is done by using the crypto or test tool to test the strength of the ciphertext of the vigenere cipher algorithm and the modification of the vigenere cipher, in this case consisting of 60 observations and the observations are tested using a sign test with the analyzed data constituting ordinal data (strong or weak).

## E. Analysis Results

The results of the analysis of observational data in this study use SPSS 21 tools through a sign test. The sign test results using SPSS 21 tools are:

Table1. Test Frequency Sign

**Frequencies**

|  |  | N |
|---|---|---|
| Y - X | Negative Differences[a] | 0 |
|  | Positive Differences[b] | 38 |
|  | Ties[c] | 22 |
|  | Total | 60 |

a. Y < X
b. Y > X
c. Y = X

**Information :**
X = Value before treatment (vigenere cipher) Y = Value after treatment (modification of vigenere cipher)
From the table above there are positive and negative signs.

Table 2. Number of Frequencies

| Sign | Frekuensi |
|---|---|
| Positif | 38 |
| Negatif | 0 |
| Ties (sama) | 22 |

On observation it can be presented as follows:

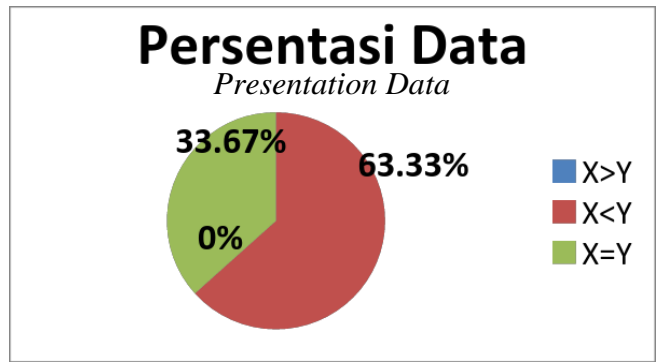

Fig8. Data Percentage Results

The level of significance of the difference between the two treatments can be determined by looking at the sig value. (2-tailed) are as follows:

Table 6: Statistical Tests

**Test Statistics[a]**

|  | Y - X |
|---|---|
| Exact Sig. (2-tailed) | .000[b] |

a. Sign Test
b. Binomial distribution used.

## F. Discussion of Analysis

The research model produces hypotheses as explained. Based on table 6 the value of sig is 0.00 and if sig <0.05 or 0.00 <0.05 then there is a significant effect with the modification of the vigenere cipher in the ciphertext arranged to not be attacked by other parties in this study through the Kasiskimethod.

From the results of testing the hypothesis, we can conclude it that they accept the proposed hypothesis, then this research supports the hypothesis that the modification of the vigenere cipher algorithm has a positive effect on the strength of the ciphertext that has been built.

## CONCLUSION

Based on the research that has been carried out, there are several conclusions:

1. The encryption and decryption process of the vigenere cipher algorithm and the modification of the vigenere cipher produce a different ciphertext even though using the same key.

2. Kasiski method uses the cryptogram of the ciphertext which is composed because of repetitive key repetition, this is the weakness of the vigenere cipher algorithm so that the key length can be determined without having to know the key used in the encryption process.

3. The key length used from the encryption process (ciphertext), not all can be known by the kaiseki method if there is no cryptogram, it is proven that the ciphertext composed of vigenere modification algorithm can influence the strength of the ciphertext that is built.

4. Modification of the vigenere cipher has a positive effect on the strength of the cipher on the vigenere cipher algorithm to make it difficult to solve the key to the Kasiski method attack by avoiding repetitive cryptograms.